\documentclass[aps, pra, twocolumn, superscriptaddress]{revtex4-2}

\usepackage{hyperref, graphicx, xcolor, amsmath, amssymb, siunitx}
\usepackage[caption=false]{subfig}

\hypersetup{colorlinks, linkcolor=teal, citecolor=blue, urlcolor=purple}

\begin{document}

\title{Magnifying quantum phase fluctuations with Cooper-pair pairing}
\author{W.\,C.~Smith}
\email[]{william.smith@ens.fr}
\author{M.~Villiers}
\author{A.~Marquet}
\affiliation{Laboratoire de Physique de l'Ecole Normale Sup\'{e}rieure, ENS, Universit\'{e} PSL, CNRS, Sorbonne Universit\'{e}, Universit\'{e} Paris-Diderot, Sorbonne Paris Cit\'{e}, Paris, France}
\affiliation{QUANTIC Team, Inria de Paris, 2 rue Simone Iff, 75012 Paris, France}
\author{J.\ Palomo}
\author{M.\,R.~Delbecq}
\author{T.~Kontos}
\affiliation{Laboratoire de Physique de l'Ecole Normale Sup\'{e}rieure, ENS, Universit\'{e} PSL, CNRS, Sorbonne Universit\'{e}, Universit\'{e} Paris-Diderot, Sorbonne Paris Cit\'{e}, Paris, France}
\author{P.~Campagne-Ibarcq}
\affiliation{QUANTIC Team, Inria de Paris, 2 rue Simone Iff, 75012 Paris, France}
\affiliation{Laboratoire de Physique de l'Ecole Normale Sup\'{e}rieure, ENS, Universit\'{e} PSL, CNRS, Sorbonne Universit\'{e}, Universit\'{e} Paris-Diderot, Sorbonne Paris Cit\'{e}, Paris, France}
\author{B.~Dou\c{c}ot}
\affiliation{Laboratoire de Physique Th\'{e}orique et Hautes Energies, Sorbonne Universit\'{e} and CNRS UMR 7589, 4 place Jussieu, 75252 Paris Cedex 05, France}
\author{Z.~Leghtas}
\email[]{zaki.leghtas@ens.fr}
\affiliation{Centre Automatique et Syst\`{e}mes, Mines-ParisTech, PSL Research University, 60 bd Saint-Michel, 75006 Paris, France}
\affiliation{Laboratoire de Physique de l'Ecole Normale Sup\'{e}rieure, ENS, Universit\'{e} PSL, CNRS, Sorbonne Universit\'{e}, Universit\'{e} Paris-Diderot, Sorbonne Paris Cit\'{e}, Paris, France}
\affiliation{QUANTIC Team, Inria de Paris, 2 rue Simone Iff, 75012 Paris, France}
\date{\today}


\begin{abstract}
Remarkably, complex assemblies of superconducting wires, electrodes, and Josephson junctions are compactly described by a handful of collective phase degrees of freedom that behave like quantum particles in a potential. Almost all these circuits operate in the regime where quantum phase fluctuations are small---the associated flux is smaller than the superconducting flux quantum---although entering the regime of large fluctuations would have profound implications for metrology and qubit protection. The difficulty arises from the apparent need for circuit impedances vastly exceeding the resistance quantum. Independently, exotic circuit elements that require Cooper pairs to form pairs in order to tunnel have been developed to encode and topologically protect quantum information. In this work we have demonstrated that pairing Cooper pairs magnifies the phase fluctuations of the circuit ground state. We measure a tenfold suppression of flux sensitivity of the first transition energy only, implying a twofold increase in the vacuum phase fluctuations, and showing that the ground state is delocalized over several Josephson wells.
\end{abstract}

\maketitle

\section{Introduction}


In superconducting circuits, patterned electrodes are arranged in seemingly endless varieties to build hardware ranging from qubits to amplifiers \cite{Clarke1988, Devoret2013}. Formally, their behavior reduces to that of a particle moving across corrugating Josephson potential wells superimposed on a parabolic background due to the circuit inductance. The finite particle mass, provided by the circuit capacitance, induces quantum fluctuations of the position-like degree of freedom---the superconducting phase \cite{Devoret1997, Girvin2014}. These circuits operate predominantly in the regime of small phase fluctuations, where the system ground state resides in a single potential well \cite{Kjaergaard2020}. What would emerge from circuits whose ground states are delocalized among multiple Josephson wells?


Computational states that are spread in phase space over multiple sites are at the root of robust and hardware-efficient encoding of quantum information. In this scheme, the delocalization is engineered to combat, at the Hamiltonian level, the detrimental effects of noise stemming from local sources. The so-called grid states of the Gottesman-Kitaev-Preskill (GKP) error correction code and the $0$-$\pi$ qubit constitute notable examples \cite{Gottesman2001, Kitaev2006}. In addition, such circuits could be tiled into topologically protected qubits \cite{Doucot2012} and employed for fault-tolerant error syndrome measurements \cite{Brooks2013, Cohen2017}, the development of quantum phase-slip elements \cite{Mooij2006}, and the observation of Bloch oscillations \cite{Piquemal2000}. Despite the wide scope of application, this regime has remained largely inaccessible.


To understand this, consider a generic superconducting qubit that consists of an inductance, a capacitance, and a Josephson junction of energies $\mathcal{E}_\mathrm{L}$, $\mathcal{E}_\mathrm{C}$, and $\mathcal{E}_\mathrm{J}$. Two parameters condition the regime of operation: $\mathcal{E}_\mathrm{J}/\mathcal{E}_\mathrm{L}$, which specifies the number of potential wells, and $\mathcal{E}_\mathrm{J}/\mathcal{E}_\mathrm{C}$, which determines the number of energy levels within a well. Here, the regime of interest corresponds to $\mathcal{E}_\mathrm{J}/\mathcal{E}_\mathrm{C}\sim 1$ and $\mathcal{E}_\mathrm{J}/\mathcal{E}_\mathrm{L}\ggg 1$, implying $\mathcal{E}_\mathrm{C}/\mathcal{E}_\mathrm{L} \ggg 1$. The latter condition corresponds to a linear circuit with characteristic impedance far exceeding the superconducting resistance quantum, which constitutes a formidable challenge. Initial progress was made using the advent of the superinductance \cite{Manucharyan2009}, now realized in various materials \cite{Hazard2019, Grunhaupt2019, PitaVidal2020, Peruzzo2020}. However, surpassing this threshold by two orders of magnitude has only recently been achieved in a feat of nanofabrication, where Josephson junction arrays were suspended in vacuum \cite{Pechenezhskiy2020}.


On the other hand, in the area of topologically protected Josephson qubits \cite{Doucot2012}, quantum information is robustly encoded in the highly degenerate ground state manifold of a fabric of circuit elements. The workhorse of these circuits is an exotic junction that forbids single Cooper-pair tunneling by Aharonov-Bohm interference, allowing only pairs of Cooper pairs to tunnel and resulting in an inherent two-fold ground state degeneracy \cite{Doucot2002, Ioffe2002.1}.


In this article, we have demonstrated that pairing Cooper pairs magnifies quantum phase fluctuations. This enhancement originates from the cancellation of single Cooper-pair tunneling, which is a significant departure (but not mutually exclusive) from the common technique of increasing the length of a kinetic inductance. We accomplish this using a generalized Josephson element, which can be tuned in situ between two operating points where the potential is $2\pi$- or $\pi$-periodic \cite{Blatter2001, Gladchenko2009}. This doubles the frequency of the corrugation and hence also the number of sites accessible to the ground state. We refer to this change of scale, resulting from a denser packing of the Josephson wells, as a magnification. In our experimental realization, we shunt this generalized Josephson element with a conventional superinductance and measure the low-lying transition energies as a function of the external magnetic flux threading the loop. We observe a tenfold suppression of the dispersion of the first excited transition energy in flux, corresponding to a twofold increase in the quantum fluctuations of the phase, which essentially washes out the effects of localization.

\section{Toy model}

\begin{figure}[t]
\centering
\subfloat{%
\includegraphics{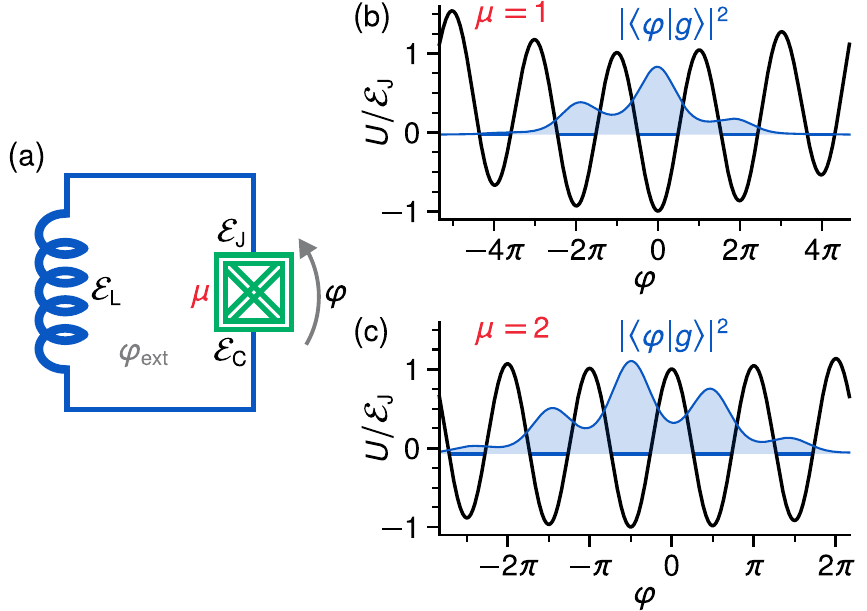}
\label{fig:simple_circuit}}
\subfloat{\label{fig:potential1}}
\subfloat{\label{fig:potential2}}
\caption{Principle of magnifying the quantum phase fluctuations. (a) Electrical circuit diagram for a generalized Josephson junction (cross-hatched box) shunted by a superinductance. This junction possesses an internal parameter $\mu=1,2,\dots$ that specifies the periodicity of the Josephson potential. (b--c) Potential and ground state probability density for the cases $\mu=1$ and $\mu=2$ at $\varphi_\mathrm{ext} = \pi/3$, showing the increase in number of Josephson wells accessible to the ground state.\label{fig:sketch}}
\end{figure}

The lumped element circuit we aim to implement (represented in Fig.\ \ref{fig:simple_circuit}) is described by the Hamiltonian
\begin{equation}
\mathcal{H}_\mu = 4 \mathcal{E}_\mathrm{C} \left( \frac{N}{\mu} \right)^2 + \frac{1}{2} \mathcal{E}_\mathrm{L} (\varphi - \varphi_\mathrm{ext})^2 + (-1)^\mu \mathcal{E}_\mathrm{J} \cos (\mu \varphi)\;.\label{eq:fluxonium}
\end{equation}
Here, $\varphi$ is the gauge-invariant superconducting phase drop across the generalized Josephson element and $N$ is the conjugate number of tunneled Cooper pairs. The external magnetic flux threading the loop is $\varphi_\mathrm{ext}$ (in dimensions of phase) and the parameter $\mu = 1,2,\dots$ specifies the multiplicity of Cooper pairs that tunnel, as can be seen in the charge basis:
\begin{equation}
\cos (\mu \varphi) = \frac{1}{2} \sum_{N=-\infty}^\infty \left( |N\rangle \langle N+\mu| + |N+\mu\rangle \langle N | \right)\:.\label{eq:tunneling}
\end{equation}
Note the occurrence of $\mu$ in the charging term in Eq.\ \ref{eq:fluxonium}, motivated by our particular implementation and to keep the generalized Josephson element plasma frequency constant. The $(-1)^\mu$ factor guarantees that the cosine potential is pinned to 1 at $\pm\pi$ for all $\mu$. In the fluxonium regime $\mathcal{E}_\mathrm{L} \ll \mathcal{E}_\mathrm{C} \lesssim \mathcal{E}_\mathrm{J}$, the ground state is confined to the lowest few potential wells \cite{Manucharyan2009}. When the period of the Josephson potential is halved by changing $\mu=1$ to $\mu=2$, the ground state delocalization is substantially magnified (see Figs.\ \ref{fig:potential1}--\ref{fig:potential2}).


To see this another way, in the rescaled coordinates $\varphi \rightarrow \varphi/\mu$ and $N \rightarrow \mu N$, the amplitude of the zero-point fluctuations of the phase across the generalized Josephson element is given by
\begin{equation}
\varphi_\mathrm{zpf} = \left(\frac{2\mathcal{E}_\mathrm{C}}{\mathcal{E}_\mathrm{L}}\mu^2\right)^{1/4}\;.\label{eq:phizpf} 
\end{equation}
Incrementing $\mu$ from one to two magnifies $\varphi_\mathrm{zpf}$ by a factor $\sqrt{2}$. Putting this into perspective, this is the enhancement that would be obtained from reducing the stray capacitance in a Josephson junction array by a factor of sixteen, allowing a fourfold increase in the total linear inductance, a strategy followed by Ref.\ \cite{Pechenezhskiy2020}. Interestingly, recasting $\sqrt{\mathcal{E}_\mathrm{C}/\mathcal{E}_\mathrm{L}}$ in Eq.\ \ref{eq:phizpf} in terms of the ratio of circuit impedance $Z_0$ to vacuum impedance $Z_\mathrm{vac}$ yields
\begin{equation}
\varphi_\mathrm{zpf} = \sqrt{8 \pi \alpha \mu \frac{Z_0}{Z_\mathrm{vac}}} \;,\label{eq:fine_structure}
\end{equation}
where $\alpha \approx 1/137$ is the fine structure constant. Regarding the shunting circuit and vacuum impedances as independent of $\mu$ indicates that here, the magnified fine structure constant $\alpha \times \mu$ for pairs of Cooper pairs appears twice as large as for Cooper pairs. This analogy provides insight into a world where charge and phase fluctuations naturally occur at the same scale. Indeed the ratio of flux fluctuations (relative to $\Phi_0$) and charge fluctuations (relative to $2e$) is on the scale of 8 times the fine structure constant \cite{Manucharyan2012.1}. In contrast, in a circuit where electrons can only move in packets of $\mu$ Cooper pairs, one finds that the ratio of flux to charge fluctuations gains a factor of $\mu$. Taking $\mu = 1/(8\alpha) \approx 17$, one would obtain a circuit where charge and flux fluctuate on similar scales.


\begin{figure}[t]
\centering
\subfloat{%
\includegraphics{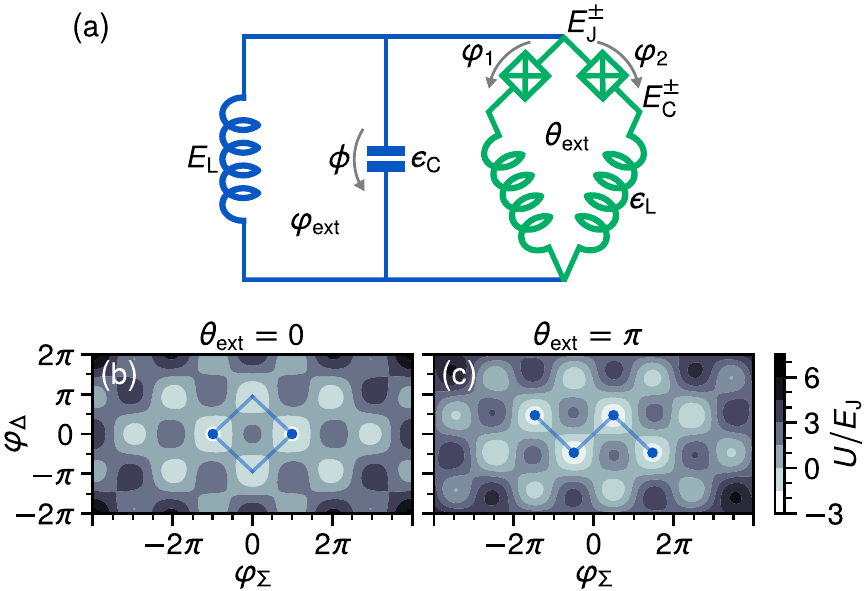}
\label{fig:circuit_reduced}}
\subfloat{\label{fig:potential_zero}}
\subfloat{\label{fig:potential_half}}
\caption{Schematic of the experimental device. (a) Electrical circuit diagram for the KITE (green) shunted by a superinductance and a capacitance (blue). Additionally, the two small Josephson junctions are allowed to have slightly differing areas, so $E_\mathrm{J}^\pm = (1 \pm \epsilon) E_\mathrm{J}$ and $E_\mathrm{C}^\pm = E_\mathrm{C} / (1 \pm \epsilon)$, with $\epsilon$ being a small asymmetry parameter. (b--c) Potential energy landscape at the external flux bias points $\varphi_\mathrm{ext} = \pi$ and $\theta_\mathrm{ext} = 0,\pi$. Here, $\varphi_\Sigma$ and $\varphi_\Delta$ are the symmetric and antisymmetric combinations of the phase drops across the two junctions of the KITE. The low-energy physics is dominated by tunneling (semi-transparent blue segments) between the lowest potential wells (solid blue dots).\label{fig:circuit}}
\end{figure}

\section{Circuit implementation}


Our implementation of the generalized Josephson element follows the proposal of Ref.\ \cite{Smith2020} (see green portion of Fig.\ \ref{fig:circuit_reduced}). A superconducting loop is composed of two nearly identical small Josephson junctions, with tunneling energy $E_\mathrm{J}$ and charging energy $E_\mathrm{C}$, in parallel and each in series with a superinductance with inductive energy $\epsilon_\mathrm{L} \ll E_\mathrm{J}$. In contrast with the rhombus \cite{Blatter2001}, we name this dipole element the Kinetic Interference coTunneling Element (KITE). Unlike the rhombus, the KITE does not contain any isolated superconducting islands and is therefore immune from offset charge noise \cite{Gladchenko2009, Bell2014}. An additional advantage of the KITE is the reduced sensitivity to flux noise within the loop owing to the large internal inductance. The KITE behaves like the generalized Josephson element depicted in Fig.\ \ref{fig:simple_circuit} with $\mu=1$ at the threaded external flux value $\theta_\mathrm{ext} = 0$ and $\mu=2$ at $\theta_\mathrm{ext} = \pi$, which we hereafter refer to as zero and half flux. While a single ridge of potential wells at zero flux leads to a spacing of $2\pi$ in the symmetric phase $\varphi_\Sigma$ between adjacent minima (see Fig.\ \ref{fig:potential_zero}), two ridges are interleaved at half flux, reducing the spacing to $\pi$ (see Fig.\ \ref{fig:potential_half} and App.\ \ref{app:reduction}) \cite{Gyenis2021}. Physically, this corresponds to the complete destructive Aharonov-Bohm interference of single Cooper-pair tunneling between the two arms of the KITE at half flux. We shunt the KITE with a superinductance (blue) of inductive energy $E_\mathrm{L}$, and its associated capacitance of charging energy $\epsilon_\mathrm{C}$. The parameter regime that we operate in is $E_\mathrm{L} < \epsilon_\mathrm{L} \ll E_\mathrm{J} \sim E_\mathrm{C}$. All superinductances are implemented with Josephson junction arrays and $\epsilon_\mathrm{C}$ is determined by the specific geometric arrangement of the connection wires.

\section{Circuit quantization}

The circuit in Fig.\ \ref{fig:circuit_reduced} has three degrees of freedom: two corresponding to the KITE and one to the shunting circuit. We obtain the Hamiltonian \cite{Devoret1997}
\begin{align}
H &= \frac{2 E_\mathrm{C}}{1 - \epsilon^2} (N_\Sigma^2 + N_\Delta^2 - 2 \epsilon N_\Sigma N_\Delta) + 4 \epsilon_\mathrm{C} n^2 + \frac{1}{2} E_\mathrm{L} \phi^2 \nonumber \\
&\qquad + \epsilon_\mathrm{L} \left(\phi - \varphi_\Sigma - \varphi_\mathrm{ext} - \tfrac{1}{2} \theta_\mathrm{ext} \right)^2 + \epsilon_\mathrm{L} \left(\varphi_\Delta - \tfrac{1}{2} \theta_\mathrm{ext}\right)^2 \nonumber \\
&\qquad - 2E_\mathrm{J} \cos \varphi_\Sigma \cos \varphi_\Delta + 2\epsilon E_\mathrm{J} \sin \varphi_\Sigma \sin \varphi_\Delta\;,\label{eq:3mode_ham}
\end{align}
where $\epsilon$ is a small junction asymmetry parameter. Due to the high degree of symmetry in the KITE, we have introduced the symmetric and antisymmetric variables $\varphi_\Sigma = \frac{1}{2}(\varphi_1 + \varphi_2)$ and $\varphi_\Delta = \frac{1}{2} (\varphi_1 - \varphi_2)$ in the above. Note also that $\{N_\Sigma, N_\Delta, n\}$ are the conjugate Cooper pair numbers to $\{\varphi_\Sigma, \varphi_\Delta, \phi\}$. Numerical diagonalization of Eq.\ \ref{eq:3mode_ham} is readily carried out in the three-mode Fock basis \cite{Smith2016}.

Our analysis focuses on the regime where $\epsilon_\mathrm{C}$ is the dominant energy scale, so we proceed by applying the Born-Oppenheimer approximation to project the $\phi$ mode into its ground state and obtain a Hamiltonian depending only on $\varphi_\Sigma$ and $\varphi_\Delta$ (see App.\ \ref{app:reduction}). Afterward, since $E_\mathrm{J}$ is the largest remaining energy scale, we adopt a two-dimensional tight-binding model. At the bias point $\theta_\mathrm{ext} = \pi$, this model can be further mapped to the one-dimensional continuous Hamiltonian
\begin{equation}
H_\pi = 4 E_\mathrm{C} \left(\frac{N}{2}\right)^2 + \frac{E_\mathrm{L} \epsilon_\mathrm{L}}{E_\mathrm{L} + 2 \epsilon_\mathrm{L}} (\varphi + \varphi_\mathrm{ext})^2 + E_\mathrm{J} \cos 2\varphi \;,\label{eq:1mode_ham_half}
\end{equation}
where $\varphi$ approximately follows the degree of freedom $\varphi_\Sigma$, which in turn approximately follows $\phi$ in the limit of $E_\mathrm{L} \ll \epsilon_\mathrm{L}$ (see App.\ \ref{app:reduction}). We also define the magnified amplitude of zero-point fluctuations of the superconducting phase \cite{Girvin2014},
\begin{equation}
\varphi_{\mathrm{zpf},\pi} = \left( 8 E_\mathrm{C} \frac{E_\mathrm{L} + 2 \epsilon_\mathrm{L}}{2 E_\mathrm{L} \epsilon_\mathrm{L}} \right)^{1/4}\;.\label{eq:phizpf_half}
\end{equation}
On the other hand, at the bias point $\theta_\mathrm{ext} = 0$, this model can be mapped to
\begin{equation}
H_0 = 4 \tilde{E}_\mathrm{C} N^2 + \frac{E_\mathrm{L} \epsilon_\mathrm{L}}{E_\mathrm{L} + 2 \epsilon_\mathrm{L}} (\varphi + \varphi_\mathrm{ext})^2 - \tilde{E}_\mathrm{J} \cos \varphi \;,\label{eq:1mode_ham_zero}
\end{equation}
where $\tilde{E}_\mathrm{J}$ and $\tilde{E}_\mathrm{C}$ are effective Josephson and charging energies that are not readily cast in terms of the other circuit parameters. Again, we define the amplitude of zero-point fluctuations of the phase
\begin{equation}
\varphi_{\mathrm{zpf},0} = \left( 2 \tilde{E}_\mathrm{C} \frac{E_\mathrm{L} + 2 \epsilon_\mathrm{L}}{2 E_\mathrm{L} \epsilon_\mathrm{L}} \right)^{1/4}\;.\label{eq:phizpf_zero}
\end{equation}
We complete the connection between our experimental circuit and the desired model by observing that Eqs.\ \ref{eq:1mode_ham_half} and \ref{eq:1mode_ham_zero} map directly onto Eq.\ \ref{eq:fluxonium} in the cases of $\mu=2$ and $\mu=1$, respectively.


\begin{table}
\centering
\begin{tabular}{c c c c c c c c}
\hline
\hline
\rule{0pt}{2.5ex} $E_\mathrm{J}/h$ & $E_\mathrm{C}/h$ & $E_\mathrm{L}/h$ & $\epsilon_\mathrm{L}/h$ & $\epsilon_\mathrm{C}/h$ & $\tilde{E}_\mathrm{J}/h$ & $\tilde{E}_\mathrm{C}/h$ & $\epsilon$ \\
\hline
5.9 & 6.6 & 0.23 & 0.36 & 2.5 & 1.7 & 1.5 & 0.03 \\
\hline
\hline
\end{tabular}
\caption{Extracted parameters for the circuit in Fig.\ \ref{fig:circuit}. The final two energy scales correspond to the effective one-dimensional Hamiltonian at $\theta_\mathrm{ext}=0$ in Eq.\ \ref{eq:1mode_ham_zero}. All energy scales are given in gigahertz.\label{tab:params}}
\end{table}

\section{Experimental realization}

\begin{figure}[t]
\centering
\subfloat{%
\includegraphics{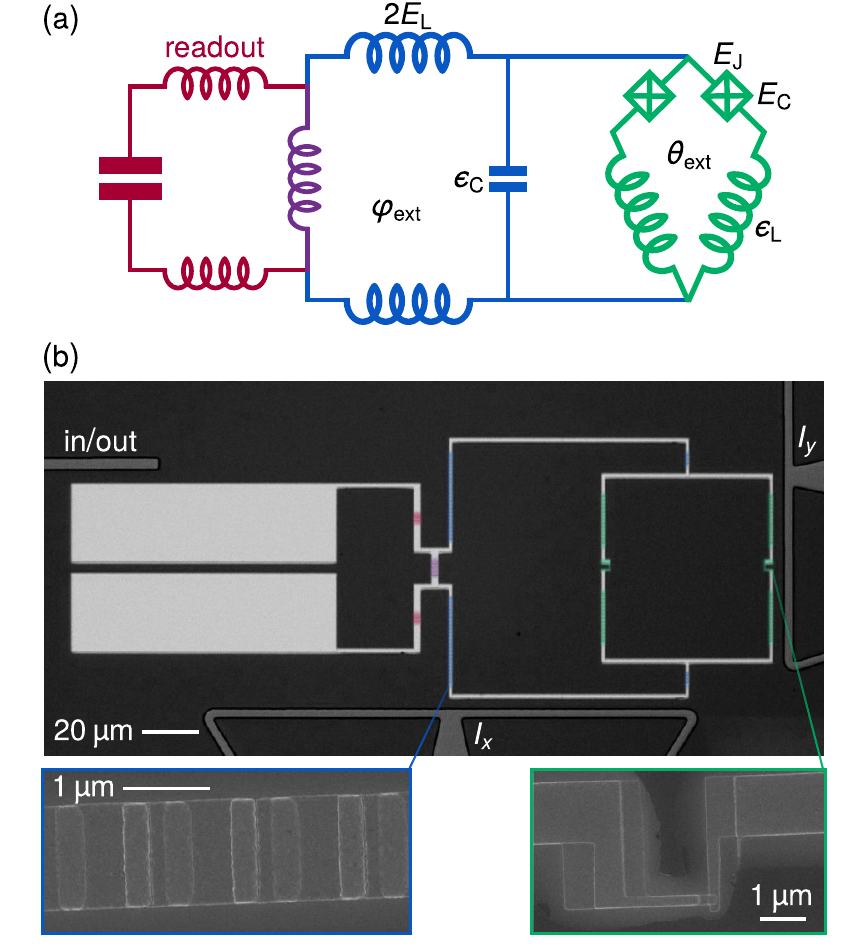}
\label{fig:full_circuit}}
\subfloat{\label{fig:sample}}
\caption{Experimental implementation. (a) Extended electrical circuit diagram for the inductively-shunted KITE including the lumped $LC$ oscillator (maroon) added for dispersive readout, which couples inductively to the circuit through a shared inductance (purple). (b) Optical micrograph of the physical device, with aluminum electrodes in light grey and niobium electrodes in dark grey. Direct currents, microwave drives, and readout signals are routed in and out of the system through two on-chip flux bias lines (right and bottom) and one weakly-coupled pin (top left). (Insets) Scanning electron microscope images of one array of large junctions [all inductances pictured in (a) are implemented similarly] and one small junction.\label{fig:device}}
\end{figure}

The circuit in Fig.\ \ref{fig:circuit_reduced} is measured through a lumped $LC$ readout resonator (maroon), which is coupled through a shared inductance (purple) as shown in Fig.\ \ref{fig:full_circuit}. We fabricate the circuit by first sputtering niobium over a silicon substrate to define the control lines (see Fig.\ \ref{fig:sample}). Then, a double-angle aluminum evaporation step produces the two small junctions and the 100 total large junctions of the KITE, as well as the 100 large junctions of the shunting superinductance. The number of superinductance junctions was chosen to keep all the self-resonant array modes above $\SI{10}{\giga\hertz}$ \cite{Masluk2012}. We extract the circuit parameters listed in Tab.\ \ref{tab:params}. The device is then measured in a dilution refrigerator at $\SI{10}{\milli\kelvin}$.


\begin{figure}
\centering
\subfloat{%
\includegraphics{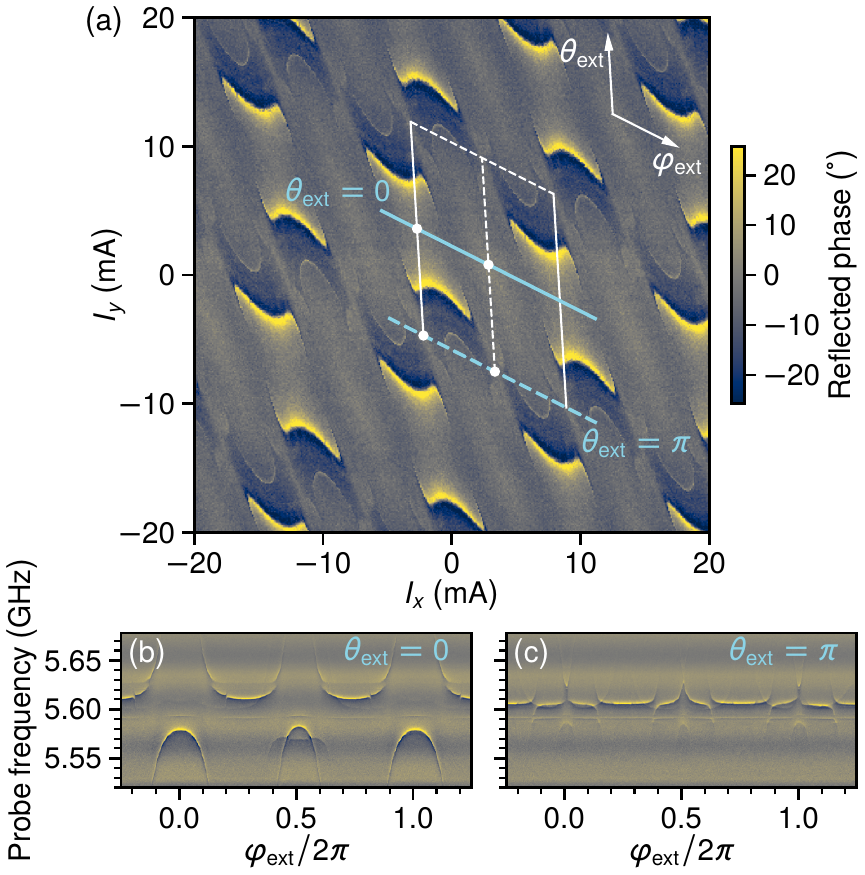}
\label{fig:fluxmap}}
\subfloat{\label{fig:antennazero}}
\subfloat{\label{fig:antennahalf}}
\caption{Device flux dependence. (a) Phase of the reflected probe signal (color) on the readout port, at the fixed probe frequency of $\SI{5.595}{\giga\hertz}$, as a function of both flux bias currents $I_x$ and $I_y$. The overlaid lines indicate the independent $\varphi_\mathrm{ext}$ and $\theta_\mathrm{ext}$ axes, while dots mark points of inversion symmetry. This phase response signals the periodic approach of the circuit transition energies in the vicinity of the readout resonance. (b--c) Phase of the reflected probe signal (color) on the readout port, as a function of probe frequency and external flux $\varphi_\mathrm{ext}$ at zero and half flux [solid and dashed blue lines in (a)], displaying periodic avoided crossings.\label{fig:flux}}
\end{figure}


We observe the phase response of the readout resonator (see Fig.\ \ref{fig:fluxmap}), which is periodically repelled by transition energies of the circuit (see Figs.\ \ref{fig:antennazero}--\ref{fig:antennahalf}), as we sweep the on-chip flux bias currents \cite{Zhang2021}. Each of the small KITE junctions contributes a cosinusoidal Josephson potential, resulting in a circuit Hamiltonian that is $2\pi$-periodic in both loop fluxes and symmetric about the points $(\varphi_\mathrm{ext}, \theta_\mathrm{ext}) = (0,0)$, $(\pi,0)$, $(0,\pi)$, and $(\pi,\pi)$ (see Eq.\ \ref{eq:3mode_ham}). These symmetries are observed in the data in Fig.\ \ref{fig:fluxmap}, where we trace the primitive cell closest to the zero-bias point (overlaid lines), which sets the scale and angle of the $\varphi_\mathrm{ext}$ and $\theta_\mathrm{ext}$ axes. The mapping between the flux bias currents $(I_x, I_y)$ and $(\varphi_\mathrm{ext}, \theta_\mathrm{ext})$ is completed by identifying the four points of inversion symmetry within the primitive cell (solid white dots) and labeling them based on the fine energy structure of the system.


\begin{figure*}
\centering
\subfloat{%
\includegraphics{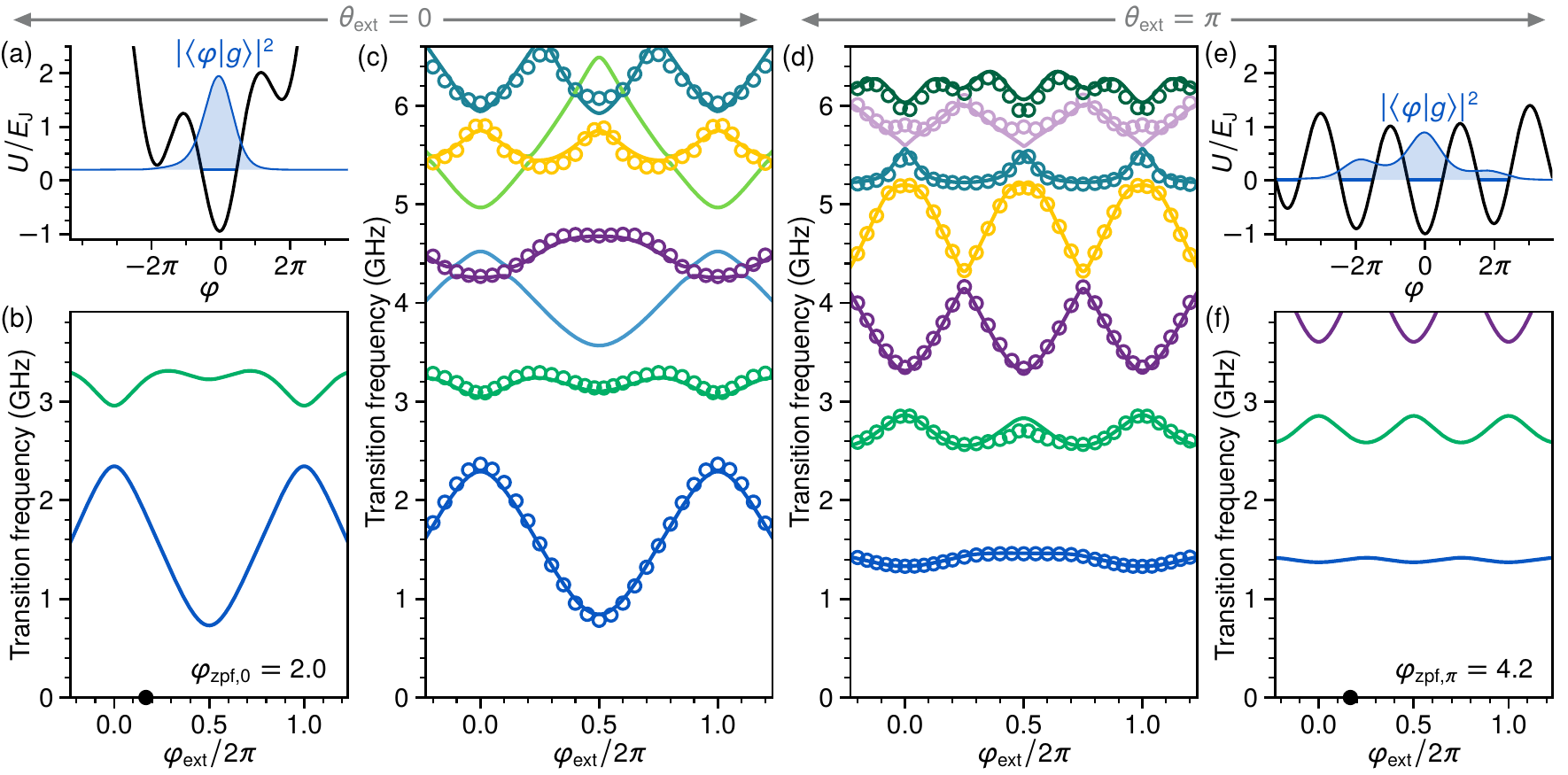}
\label{fig:wavefunctionzero}}
\subfloat{\label{fig:phizpfzero}}
\subfloat{\label{fig:levelszero}}
\subfloat{\label{fig:levelshalf}}
\subfloat{\label{fig:wavefunctionhalf}}
\subfloat{\label{fig:phizpfhalf}}
\caption{Magnified quantum phase fluctuations. (a,e) Potential energy landscape and ground state probability density at $\varphi_\mathrm{ext} = \pi/3$ (black dot in plots beneath) for the cases of zero flux $\theta_\mathrm{ext} = 0$ and half flux $\theta_\mathrm{ext} = \pi$. (b,f) Lowest-lying transition frequencies of the effective single-mode Hamiltonians at zero and half flux in Eqs.\ \ref{eq:1mode_ham_zero} and \ref{eq:1mode_ham_half} with parameters in Tab.\ \ref{tab:params}, showing an increase in the amplitude of the zero-point fluctuations of the phase from $\varphi_{\mathrm{zpf},0} = 2.0$ to $\varphi_{\mathrm{zpf},\pi} = 4.2$. (c--d) Two-tone measurements of the transition frequencies from the ground state (open circles) and corresponding fitted values (solid curves) obtained from the three-mode circuit. The light blue and light green transitions were not visible in spectroscopy due to their strong decoupling from the readout resonator. The essential feature is the remarkable suppression of flux dispersion of the first excited state transition when $\theta_\mathrm{ext}$ is stepped from $0$ to $\pi$.\label{fig:levels}}
\end{figure*}

\section{Transition energies}

We access the transition energies of the circuit in a two-tone spectroscopy experiment at $\theta_\mathrm{ext} = 0, \pi$ (see Figs.\ \ref{fig:levelszero}--\ref{fig:levelshalf}). For each external flux value $\varphi_\mathrm{ext}$, we calibrate the frequency of a readout tone and measure its reflected phase response as a function of the frequency of a probe tone fed into the $I_x$ port. In order to fully constrain the circuit Hamiltonian identification, we report the fitted center frequencies of all persistent spectral lines visible below $\SI{6.6}{\giga\hertz}$. We fit the twelve measured transition energies (five at zero flux and seven at half flux) to the eigenspectrum of the Hamiltonian in Eq.\ \ref{eq:3mode_ham}, parametrized by $\{E_\mathrm{J}, E_\mathrm{C}, E_\mathrm{L}, \epsilon_\mathrm{L}\}$ (see App.\ \ref{app:fit}). The remarkable agreement between data and theory is a powerful demonstration that this relatively complex assembly---involving two loops and 215 Josephson junctions---is well captured by a three-mode model with only four independent parameters.


The most essential feature of this spectrum is the drastic reduction of the flux dispersion of the first excited state transition, which spans $\SI{1.6}{\giga\hertz}$ at zero flux and only $\SI{140}{\mega\hertz}$ at half flux (blue circles in Figs.\ \ref{fig:levelszero}--\ref{fig:levelshalf}). Since the higher transitions exhibit significant flux dependence, this reduction is a direct signature of the ground and first excited state delocalization over additional Josephson wells. Indeed, as the external flux is increased, the cosinusoidal corrugation slides over the parabolic background. A particle localized in a single Josephson well---known as a fluxon state---would be carried upward in energy until it tunnels to the trailing descending fluxon state, resulting in a large flux dispersion. On the other hand, a delocalized particle simultaneously occupies ascending and descending fluxon states, washing out the flux dependence of the energy.

An additional striking feature of Fig.\ \ref{fig:levelshalf} is the dominant $\pi$-periodicity of the energy spectrum at $\theta_\mathrm{ext} = \pi$, in stark contrast with the usual $2\pi$-periodicity of a Josephson potential, as observed in Fig.\ \ref{fig:levelszero}. This effect is slightly diminished by residual asymmetry in the areas of the two small KITE junctions, which we find to be around $\SI{6}{\percent}$. Finally, note the two transitions at zero flux (light blue and light green lines in Fig.\ \ref{fig:levelszero}) predicted by theory but not observed in the experiment. This is explained by the vanishingly small dispersive coupling of these transitions to the readout resonator, as predicted by circuit quantization \footnote{Although we have neglected the readout mode in Eq.\ \ref{eq:3mode_ham}, we note that it directly couples to the $\phi$ mode exclusively. Since the $\phi$ variable hybridizes significantly with $\varphi_\Sigma$, but not $\varphi_\Delta$, we expect only a subset of the eigenstates to dispersively shift the readout resonance at external flux values where $\varphi_\Sigma$ and $\varphi_\Delta$ are roughly independent. This is the case at $\theta_\mathrm{ext} = 0$ but not $\theta_\mathrm{ext} = \pi$.}.


Although a three-mode model was necessary to capture the complete energy spectrum up to $\SI{6.6}{\GHz}$, the first few transition energies are well described by the emergent low-energy Hamiltonians of Eqs.\ \ref{eq:1mode_ham_zero} and \ref{eq:1mode_ham_half} with parameters listed in Tab.\ \ref{tab:params}, as shown in Figs.\ \ref{fig:phizpfzero} and \ref{fig:phizpfhalf}. Note that in the experiment we observe a magnification factor of $\varphi_{\mathrm{zpf},\pi}/\varphi_{\mathrm{zpf},0} = 2.1$, which is larger than the factor of $\sqrt{2}$ predicted by Eq.\ \ref{eq:phizpf} because the effective behavior of our circuit at $\theta_\mathrm{ext} = 0$ involves double-phase-slip processes (see App.\ \ref{app:reduction}). We find $\varphi_{\mathrm{zpf},\pi} = 4.2\pm 0.2$, which is strictly larger than $\pi$, therefore unambiguously placing our system in the regime of ground state delocalization.

\section{Outlook}


An ongoing quest in quantum information science is to build a circuit that hosts a degenerate ground state manifold that encodes and protects quantum states. A particularly appealing approach is to combine a generalized Josephson element with the so-called quantum phase-slip element \cite{Mooij2006}, the dual of the Josephson junction \cite{Le2019}. Such a circuit would encode information in the form of GKP states \cite{Gottesman2001} that are protected at the Hamiltonian level, circumventing the need for error syndrome measurements and feedback. The key technological innovation of our experiment---the KITE---is an implementation of such a generalized Josephson element. This circuit would then need to be combined with additional parts that effectively play the role of a phase-slip element. We are then confronted to the usual challenge in quantum error correction, where adding complexity to a circuit may have the virtue of hosting a protected manifold of states, but comes at the cost of adding error channels. Examples are charge noise in each added island, and flux noise in each added loop. Inventing a circuit where the protection overcomes all these additional sources of errors is an active area of research \cite{Groszkowski2018}.

\section{Conclusion}

In conclusion, we have built a superconducting circuit whose ground state wavefunction spreads across multiple Josephson potential wells by packing them closer together via Cooper-pair pairing. We refer to this process as magnification of the phase fluctuations, and quantify it by an amplitude $\varphi_{\mathrm{zpf},\pi} = 4.2 \pm 0.2$. We find it intriguing that the origin of this phenomenon does not rely on the inertia (kinetic inductance) of charge carriers, but by forcing them to tunnel in pairs. We envision incorporating a quantum phase-slip element into our circuit to implement a protected qubit, observe the quantum dynamics of nonlocal operators, and perform fault-tolerant error syndrome measurements \cite{Brooks2013, Cohen2017}.


\section{Acknowledgments}

We thank Rapha\"{e}l Lescanne and Mazyar Mirrahimi for fruitful discussions and Lincoln Labs for providing a Josephson Traveling-Wave Parametric Amplifier. The devices were fabricated within the consortium Salle Blanche Paris Centre. This work was supported by the QuantERA grant QuCOS, by ANR 19-QUAN-0006-04. Z.L.\ acknowledges support from ANR project ENDURANCE, and EMERGENCES grant ENDURANCE of Ville de Paris. This work has been supported by the Paris \^{I}le-de-France Region in the framework of DIM SIRTEQ. This project has received funding from the European Research Council (ERC) under the European Union’s Horizon 2020 research and innovation programme (grant agreement No.\ 851740).


\section{Author contributions}

W.C.S.\ conceived the experiment and measured the device. W.C.S., assisted by A.M., designed the sample. Z.L.\ fabricated the sample. W.C.S, Z.L, B.D., and P.C.-I.\ analyzed the data. M.V., J.P., T.K., and M.R.D.\ provided experimental support. W.C.S.\ and Z.L.\ co-wrote the manuscript with input from all authors.

\appendix

\section{Device fabrication}

The circuit is fabricated on a $\SI{280}{\micro\meter}$-thick wafer of intrinsic silicon (resistivity exceeding $\SI{10000}{\ohm\centi\meter}$). The silicon wafer is cleaned in solvents and loaded in a sputtering system. After one night of pumping, we start with an argon milling cleaning step, and proceed to sputter $\SI{120}{\nano\meter}$ of niobium onto the chip. We spin optical resist (S1805), and pattern the large features (control lines) with a laser writer. After development (MF319), we etch the sample in SF6 with a $\SI{20}{\second}$ overetch, followed by lift off (acetone at $\SI{50}{\celsius}$). We then spin a bilayer of MAA (EL11) and PMMA (A6). The entire circuit (KITE, inductive shunt, and readout resonator) is patterned in an e-beam lithography step. The development takes place in a 3:1 IPA/water solution at $\SI{6}{\celsius}$ for $\SI{90}{\second}$, followed by $\SI{10}{\second}$ in IPA. The chip is then loaded in an e-beam evaporator. We start with a thorough argon ion milling for $\SI{2}{\minute}$ with the sample at the angles of $\SI{0}{\degree}$ and $\SI{45}{\degree}$, which serves the purpose of cleaning the substrate. We then evaporate $\SI{35}{\nano\meter}$ and $\SI{70}{\nano\meter}$ of aluminum, at $\SI{0}{\degree}$ and $\SI{45}{\degree}$ angles, separated by an oxidation step of $\SI{200}{\milli\bar}$ of pure oxygen for $\SI{10}{\minute}$.

The Josephson junctions are all fabricated from Al/AlOx/Al in a single evaporation step, using the Dolan bridge method with an e-beam base dose of $\SI{283}{\micro\coulomb\per\square\centi\meter}$ at an acceleration voltage of $\SI{20}{\kilo\volt}$, and come in three types. (i) Two small junctions in the KITE of area $\SI{0.075}{\square\micro\meter}$, made in a cross geometry, patterned with a dose factor of $1.6$, and resulting in an inductance per junction of $\SI{52}{\nano\henry}$. (ii) A total of 100 large array junctions within the KITE loop and a total of 100 unshared shunting array junctions, of area $\SI{0.49}{\square\micro\meter}$, patterned with a dose factor of $1.8$, and resulting in an inductance per junction of $\SI{6.8}{\nano\henry}$. (iii) Thirteen larger array junctions that form the readout resonator inductance (seven of which are shared) of area $\SI{1.10}{\square\micro\meter}$, patterned with a dose factor of $1.5$, and resulting in an inductance per junction of $\SI{3.45}{\nano\henry}$.

The chip was subsequently baked at $\SI{115}{\celsius}$ for $\SI{1}{\minute}$, glued with PMMA onto a PCB, wire-bonded and mounted into a sample holder. The device was then thermally anchored to the base plate of a Bluefors dilution refrigerator, surrounded by three concentric cans for magnetic and infrared shielding (outer: cryoperm, middle: aluminum, inner: copper). An optical micrograph of the circuit and SEM images of some junctions are shown in Fig.\ \ref{fig:sample}.

\section{Model reduction\label{app:reduction}}

In this section, we describe the mathematical procedure for reducing the three-mode Hamiltonian in Eq.\ \ref{eq:3mode_ham} to an effective one-mode Hamiltonian at the KITE bias values $\theta_\mathrm{ext} = \pi,0$. The main purpose of these approximations is to extract an effective amplitude of quantum phase fluctuations, and thereby benchmark the extent of observed ground state delocalization.

\subsection{Born-Oppenheimer approximation}

We first define the energy scales of the circuit according to
\begin{align}
E_\mathrm{C} &= \frac{e^2}{2 C_\mathrm{J}} & \epsilon_\mathrm{C} &= \frac{e^2}{2 C} & E_\mathrm{J} &= \frac{\phi_0^2}{L_\mathrm{J}} & E_\mathrm{L} &= \frac{\phi_0^2}{L} & \epsilon_\mathrm{L} &= \frac{\phi_0^2}{\ell}\;,
\end{align}
where $\phi_0 = \hbar / 2e$ is the reduced magnetic flux quantum, and we place ourselves in the regime where $E_\mathrm{L} < \epsilon_\mathrm{L} \ll E_\mathrm{J} \sim E_\mathrm{C}$ and $\epsilon \ll 1$ in Eq.\ \ref{eq:3mode_ham}. Furthermore, we are chiefly interested in the limit where $\epsilon_\mathrm{C}$ is the dominant energy scale, i.e.\ the limit of small shunt capacitance. While we did not strictly achieve this experimentally due to our specific circuit geometry, this limit conveniently reveals the relevant underlying physics of the circuit. In Born-Oppenheimer theory, one considers a system with a collection of high-frequency electron-like degrees of freedom and a collection of low-frequency nucleus-like degrees of freedom. To obtain an approximate Hamiltonian for the ``nuclei,'' one freezes the ``electrons'' in their quantum ground state, which is found by fixing the nuclear variables to their classical values. In our system, we regard the $\phi$ variable as high-frequency and treat the $\varphi_\Sigma$ and $\varphi_\Delta$ variables as low-frequency. Fixing $\varphi_\Sigma$ and $\varphi_\Delta$ to their classical values and setting $N_\Sigma$ and $N_\Delta$ to zero, Eq.\ \ref{eq:3mode_ham} reduces to
\begin{equation}
H_\mathrm{e} = 4 \epsilon_\mathrm{C} n^2 + \tfrac{1}{2} E_\mathrm{L} \phi^2 + \epsilon_\mathrm{L} \left(\phi - \varphi_\Sigma - \varphi_\mathrm{ext} - \tfrac{1}{2} \theta_\mathrm{ext} \right)^2\;.
\end{equation}
Note that, in this equation, $\phi$ and $n$ are operators while $\varphi_\Sigma$ is a parameter. This Hamiltonian simply describes a harmonic oscillator whose equilibrium position is $\langle \phi \rangle = \frac{2 \epsilon_\mathrm{L}}{E_\mathrm{L} + 2 \epsilon_\mathrm{L}} (\varphi_\Sigma + \varphi_\mathrm{ext} + \frac{1}{2} \theta_\mathrm{ext})$. In the limit that $E_\mathrm{L} \ll \epsilon_\mathrm{L}$, we observe that the dynamics of $\phi$ rigidly follow those of $\varphi_\Sigma$. We can hence readily identify the energy levels of $H_\mathrm{e}$ as
\begin{align}
E_{\mathrm{e}, m} &= \sqrt{8 (E_\mathrm{L} + 2 \epsilon_\mathrm{L}) \epsilon_\mathrm{C}} \left(m + \tfrac{1}{2} \right) \nonumber \\
&\qquad + \frac{E_\mathrm{L} \epsilon_\mathrm{L}}{E_\mathrm{L} + 2 \epsilon_\mathrm{L}} \left(\varphi_\Sigma + \varphi_\mathrm{ext} + \tfrac{1}{2} \theta_\mathrm{ext}\right)^2\;.
\end{align}
We may then plug this into Eq.\ \ref{eq:3mode_ham} to obtain the effective low-energy Hamiltonian
\begin{align}
H_\mathrm{n} &= \frac{2 E_\mathrm{C}}{1 - \epsilon^2} (N_\Sigma^2 + N_\Delta^2 - 2 \epsilon N_\Sigma N_\Delta) \nonumber \\
& + \frac{E_\mathrm{L} \epsilon_\mathrm{L}}{E_\mathrm{L} + 2 \epsilon_\mathrm{L}} \left(\varphi_\Sigma + \varphi_\mathrm{ext} + \tfrac{1}{2} \theta_\mathrm{ext}\right)^2 + \epsilon_\mathrm{L} \left(\varphi_\Delta - \tfrac{1}{2} \theta_\mathrm{ext}\right)^2 \nonumber \\
& - 2E_\mathrm{J} \cos \varphi_\Sigma \cos \varphi_\Delta + 2\epsilon E_\mathrm{J} \sin \varphi_\Sigma \sin \varphi_\Delta \nonumber \\
& + \sqrt{8 (E_\mathrm{L} + 2 \epsilon_\mathrm{L}) \epsilon_\mathrm{C}} \left(m + \tfrac{1}{2} \right)\;.\label{eq:nuclear_ham}
\end{align}
Recall that here, $\varphi_\Sigma$ and $\varphi_\Delta$ have been restored as operators (as well as $N_\Sigma$ and $N_\Delta$), while $\phi$ has been reduced to the discrete parameter $m=0,1,\dots$ (playing the role of an orbital index) while also modifying the energy landscape of $\varphi_\Sigma$. Note that the solutions to $H_\mathrm{n}$ can only be expected to resemble those to the full Hamiltonian for energies smaller than $\sqrt{8 (E_\mathrm{L} + 2 \epsilon_\mathrm{L}) \epsilon_\mathrm{C}}$. For our experimental parameters, this cutoff corresponds to a frequency of $\SI{4.35}{\giga\hertz}$, which makes this analysis relevant for the first two transitions only. Furthermore, the inductive energy for $\varphi_\Sigma$ corresponds to that for a series arrangement of inductances $L$ and $\ell/2$ (a parallel arrangement of two inductances $\ell$).

\subsection{Projection at \boldmath$\theta_\mathrm{ext} = \pi$}

Restricting our attention to the case of symmetric junctions, we consider Eq.\ \ref{eq:nuclear_ham} for $m=0$,
\begin{align}
H_{\mathrm{n},\pi} &= 2 E_\mathrm{C} (N_\Sigma^2 + N_\Delta^2) + \frac{E_\mathrm{L} \epsilon_\mathrm{L}}{E_\mathrm{L} + 2 \epsilon_\mathrm{L}} (\varphi_\Sigma + \varphi_\mathrm{ext})^2 + \epsilon_\mathrm{L} \varphi_\Delta^2 \nonumber \\
&\qquad + 2 E_\mathrm{J} \sin \varphi_\Sigma \sin \varphi_\Delta\;.\label{eq:2mode_half}
\end{align}
Since $E_\mathrm{J}$ is the dominant energy scale, we first consider the locations at which the Josephson tunneling energy is minimized: $\varphi_\Sigma = (m_\Sigma + \frac{1}{2}) \pi$ and $\varphi_\Delta = (m_\Delta + \frac{1}{2}) \pi$ where $m_\Sigma$ and $m_\Delta$ are integers with the same parity. Due to their small energy scales, the linearly inductive terms only contribute a broad parabolic confinement centered at $\varphi_\Sigma = -\varphi_\mathrm{ext}$ and $\varphi_\Delta = 0$. It is then natural to introduce a two-dimensional tight-binding model with individual sites corresponding to the different Josephson wells. To this end, we arrive at the Hamiltonian
\begin{widetext}
\begin{align}
H_{\mathrm{n},\pi}^\text{tight-binding,2D} &= \sum_{\{m_\Sigma + m_\Delta \, \text{even}\}} \left[ \frac{E_\mathrm{L} \epsilon_\mathrm{L}}{E_\mathrm{L} + 2 \epsilon_\mathrm{L}} (m_\Sigma \pi + \tfrac{\pi}{2} + \varphi_\mathrm{ext})^2 + \epsilon_\mathrm{L} (m_\Delta \pi + \tfrac{\pi}{2})^2 \right] |m_\Sigma, m_\Delta \rangle \langle m_\Sigma, m_\Delta | \nonumber \\
&\qquad - \sum_{\{m_\Sigma + m_\Delta \, \text{even}\}} \frac{1}{2} \Gamma \left( |m_\Sigma, m_\Delta \rangle \langle m_\Sigma \pm 1, m_\Delta \pm 1| + |m_\Sigma \pm 1, m_\Delta \pm 1 \rangle \langle m_\Sigma, m_\Delta | \right) \;,\label{eq:tight_half_2d}
\end{align}
\end{widetext}
where the index $\{m_\Sigma + m_\Delta \, \text{even}\}$ indicates summation over only the lowest potential minima and the hopping energy $\Gamma$ corresponds to the nearest-neighbor tunneling rate for the lowest wells. The factor of $1/2$ accounts for double counting and the hopping rate coincides with the standard result for the phase-slip rate across a single Josephson junction of tunneling energy $E_\mathrm{J}$ and charging energy $E_\mathrm{C}$ \cite{Matveev2002},
\begin{equation}
\Gamma = \frac{4}{\sqrt{\pi}} (8 E_\mathrm{J}^3 E_\mathrm{C})^{1/4} \exp \left( -\sqrt{\frac{8 E_\mathrm{J}}{E_\mathrm{C}}}\right)\;.
\end{equation}
Indeed, this is the phase-slip rate across a single KITE junction because $E_\mathrm{L} < \epsilon_\mathrm{L} \ll E_\mathrm{J}$. Note that we have neglected the effect of double phase slips, that is, transitions $m_{\Sigma,\Delta} \rightarrow m_{\Sigma,\Delta} \pm 2$ with $m_{\Delta,\Sigma}$ unchanged.

Inspection of Eq.\ \ref{eq:tight_half_2d} shows that, at all values of the external flux $\varphi_\mathrm{ext}$, there are essentially two degenerate lowest rungs of potential wells at $m_\Delta = -1, 0$. This feature makes it possible to draw a zig-zag path through the $\varphi_\Sigma \varphi_\Delta$-plane that connects all the relevant low-lying potential minima (see Fig.\ \ref{fig:potential_half}). To this end, we introduce a new tight-binding site index $s$ and write the one-dimensional Hamiltonian
\begin{align}
H_{\mathrm{n},\pi}^\text{tight-binding,1D} &= \sum_{\{s\}} \frac{E_\mathrm{L} \epsilon_\mathrm{L}}{E_\mathrm{L} + 2 \epsilon_\mathrm{L}} (s \pi + \tfrac{\pi}{2} + \varphi_\mathrm{ext})^2 |s \rangle \langle s| \nonumber \\
&\qquad - \sum_{\{s\}} \frac{1}{2} \Gamma \left(|s\rangle \langle s+1 | + |s+1 \rangle \langle s|\right) \;,\label{eq:tight_half_1d}
\end{align}
where the index $\{ s\}$ again indicates summation over only the lowest potential minima. In the above, the most important feature is that neighboring sites are separated by a distance of $\pi$, rather than the conventional $2\pi$. It is interesting to note that $s$ in Eq.\ \ref{eq:tight_half_1d} plays the role of a fluxon index, in the sense that transitions between neighboring sites involve a single fluxon tunneling in or out of the KITE. Moreover, this Hamiltonian is formally identical to that of the usual Cooper pair box under the interchange of Cooper pair number with fluxon occupation. We therefore conclude that the regime of ground state delocalization---where the tight-binding approximation holds---is also the regime where the circuit begins to behave like a quantum phase-slip element \cite{Mooij2006}.

We then observe that Eq.\ \ref{eq:tight_half_1d} is also obtained from the continuous Hamiltonian in Eq.\ \ref{eq:1mode_ham_half},
%
%
where $\varphi$ is the continuous analogue of $s$, and so follows the degree of freedom $\varphi_\Sigma$. Due to the conjugacy between $\varphi$ and $N$, as well as the correspondence from the previous section between $\phi$ and $\varphi_\Sigma$, we associate $N$ with the number of Cooper pairs that have tunneled across the KITE. This model must be treated with caution; its spectrum is only expected to resemble that of Eq.\ \ref{eq:tight_half_1d} for the lowest-energy eigenstates and there are high-energy eigenstates of Eq.\ \ref{eq:1mode_ham_half} that have no counterpart in Eq.\ \ref{eq:tight_half_1d}. As such, we present the eigenspectrum obtained by numerical diagonalization of $H_\pi$ in Fig.\ \ref{fig:phizpfhalf} with no fitted parameters. Finally, in order to benchmark the ground state delocalization, we now define the effective amplitude of zero-point fluctuations of the superconducting phase \cite{Girvin2014} in Eq.\ \ref{eq:phizpf_half}.
%

\subsection{Projection at \boldmath$\theta_\mathrm{ext} = 0$}

Following the same procedure as in the previous section, we start with the Hamiltonian
\begin{align}
H_{\mathrm{n},0} &= 2 E_\mathrm{C} (N_\Sigma^2 + N_\Delta^2) + \frac{E_\mathrm{L} \epsilon_\mathrm{L}}{E_\mathrm{L} + 2 \epsilon_\mathrm{L}} (\varphi_\Sigma + \varphi_\mathrm{ext})^2 + \epsilon_\mathrm{L} \varphi_\Delta^2 \nonumber \\
&\qquad - 2 E_\mathrm{J} \cos \varphi_\Sigma \cos \varphi_\Delta\label{eq:2mode_zero}
\end{align}
and observe that the relevant minima are located at $\varphi_\Sigma = m_\Sigma \pi$ and $\varphi_\Delta = m_\Delta \pi$ where $m_\Sigma$ and $m_\Delta$ are again integers with the same parity. Here, we arrive the tight-binding Hamiltonian
\begin{widetext}
\begin{align}
H_{\mathrm{n},0}^\text{tight-binding,2D} &= \sum_{\{m_\Sigma + m_\Delta \, \text{even}\}} \left[ \frac{E_\mathrm{L} \epsilon_\mathrm{L}}{E_\mathrm{L} + 2 \epsilon_\mathrm{L}} (m_\Sigma \pi + \varphi_\mathrm{ext})^2 + \epsilon_\mathrm{L} (m_\Delta \pi)^2 \right] |m_\Sigma, m_\Delta \rangle \langle m_\Sigma, m_\Delta | \nonumber \\
&\qquad - \sum_{\{m_\Sigma + m_\Delta \, \text{even}\}} \frac{1}{2} \Gamma \left( |m_\Sigma, m_\Delta \rangle \langle m_\Sigma \pm 1, m_\Delta \pm 1| + |m_\Sigma \pm 1, m_\Delta \pm 1 \rangle \langle m_\Sigma, m_\Delta | \right) \;,\label{eq:tight_zero_2d}
\end{align}
\end{widetext}
which crucially differs from Eq.\ \ref{eq:tight_half_2d} in the $\pi/2$ offsetted equilibrium positions in the first line.

Examination of Eq.\ \ref{eq:tight_zero_2d} reveals that, at all values of the external flux $\varphi_\mathrm{ext}$, there is a lowest rung of potential wells at $m_\Delta = 0$ and two degenerate next-lowest rungs at $m_\Delta = \pm 1$. Consequently, it is nontrivial to draw a one-dimensional path through the $\varphi_\Sigma \varphi_\Delta$-plane onto which to adequately project the dynamics, as we did at $\theta_\mathrm{ext} = \pi$. Nonetheless, in order to model the lowest energy transitions, we introduce the phenomenological one-dimensional Hamiltonian in Eq.\ \ref{eq:1mode_ham_zero}.
%
%
The energy spectrum obtained by numerical diagonalization of $H_0$, with parameters $\tilde{E}_\mathrm{J}/h = \SI{1.66}{\giga\hertz}$ and $\tilde{E}_\mathrm{C}/h = \SI{1.48}{\giga\hertz}$ determined by a fit to the data, is shown in Fig.\ \ref{fig:phizpfzero}. Here, $\varphi$ is a collective degree of freedom that experiences the same parabolic confinement as the $\varphi_\Sigma$ variable in Eq.\ \ref{eq:2mode_zero}. As above, we define the effective amplitude of zero-point fluctuations of the phase in Eq.\ \ref{eq:phizpf_zero}.
%

\section{Spectrum fitting\label{app:fit}}

\begin{figure}
\centering
\includegraphics{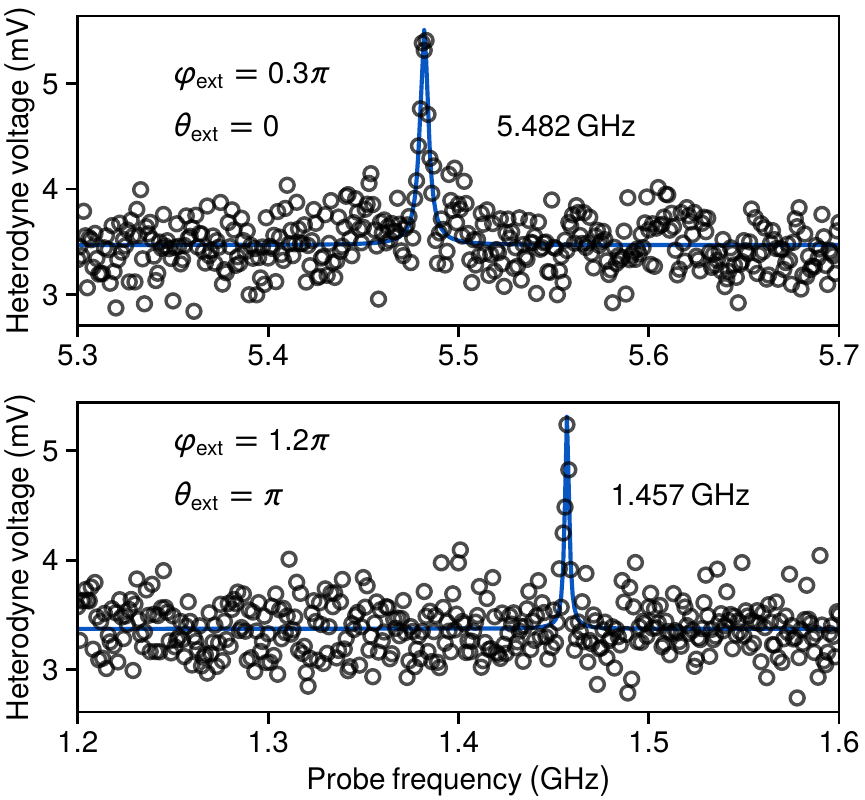}
\caption{Typical two-tone spectroscopy measurements, of the transition from the ground to fourth excited state at zero flux (top plot) and of that to the first excited state at half flux (bottom plot). Fitted Lorentzians are shown in blue.\label{fig:spec}}
\end{figure}

We fit the twelve measured transition frequencies shown in Figs.\ \ref{fig:levelszero}--\ref{fig:levelshalf}, and extracted from spectroscopy traces like those in Fig.\ \ref{fig:spec}, to the eigenspectrum of Eq.\ \ref{eq:3mode_ham}. In total this Hamiltonian has six parameters: $E_\mathrm{J}$, $E_\mathrm{C}$, $E_\mathrm{L}$, $\epsilon_\mathrm{L}$, $\epsilon_\mathrm{C}$, and $\epsilon$. We initialize our guess of $E_\mathrm{J}, E_\mathrm{L}$, and $\epsilon_\mathrm{L}$ using normal resistance measurements and the Ambegaokar-Baratoff formula, and start by setting $\epsilon$ to zero. Then, assuming a plasma frequency of $\SI{17.6}{\giga\hertz}$, consistent with previous measurements in our laboratory at similar oxidation parameters, we infer $E_\mathrm{C}$. In order to determine $\epsilon_\mathrm{C}$, which is mainly due to the capacitance of the metallic lines in the circuit, we run a finite-element simulation (Ansys HFSS) of the circuit, with all array junctions replaced by linear inductances and the two small junctions of the KITE replaced by capacitances. The simulation provides the frequency of a mode resembling the $\varphi_\Sigma$ mode. We then numerically find $\epsilon_\mathrm{C}$---in a lumped element circuit corresponding to the HFSS model---that matches the simulated frequency, ultimately finding $\epsilon_\mathrm{C}= \SI{2.5}{\giga\hertz}$, which corresponds to a shunt capacitance of $C = \SI{7.7}{\femto\farad}$. 

\begin{figure}
\centering
\includegraphics{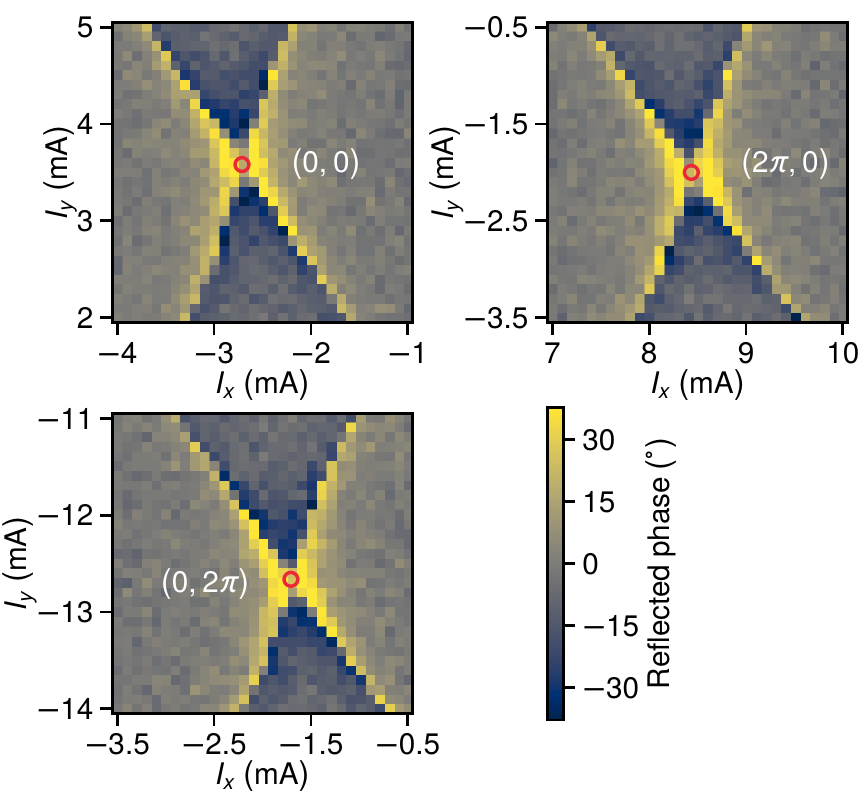}
\caption{Typical external flux calibration data. Phase of the reflected probe signal (color) on the readout port, at the fixed probe frequency of $\SI{5.580}{\giga\hertz}$, as a function of both flux bias currents $I_x$ and $I_y$. The three panels correspond to zoomed-in portions of Fig.\ \ref{fig:fluxmap}, about the inversion-symmetric points $(\varphi_\mathrm{ext}, \theta_\mathrm{ext}) = (0,0)$, $(2\pi, 0)$, and $(0, 2\pi)$. Red circles indicate symmetry points extracted using image inversion convolution, and are used to fully determine the mapping from $(I_x, I_y)$ onto $(\varphi_\mathrm{ext}, \theta_\mathrm{ext})$.\label{fig:calib}}
\end{figure}

We use all these values to initialize a least-squares fit of the measured transition frequencies to fine tune the four independent parameters $E_\mathrm{J}$, $E_\mathrm{C}$, $E_\mathrm{L}$, and $\epsilon_\mathrm{L}$. Note that, as per the discussion in the main text, we do not expect every energy transition to be easily visible in our spectroscopic measurement setup through the readout mode, and therefore appropriate fitting requires a filtering of different energy levels. Indeed, the readout mode couples to the circuit through $\varphi_\Sigma$, and therefore in the fit routine, we neglect the transition to a given state $|i\rangle$ when $|\langle i | \varphi_\Sigma | g \rangle| \ll |\langle i | \varphi_\Delta | g \rangle|$. Finally, we improve the fit by ramping up the asymmetry parameter to $\epsilon = 0.03$, which corresponds to $\SI{6}{\percent}$ relative asymmetry.

\section{External flux calibration}

We have observed that the external magnetic flux drifts on the order of $0.02 \Phi_0$, where $\Phi_0 = h / 2e$ is the magnetic flux quantum, on a timescale of $\SI{24}{\hour}$. Therefore, we calibrate the external magnetic flux axes $\varphi_\mathrm{ext}$ and $\theta_\mathrm{ext}$ daily. To this end, we monitor the readout resonator response as a function of $I_x$ and $I_y$ over a narrow range around three symmetry points, as shown in Fig.\ \ref{fig:calib}.

\section{Coherence times}

\begin{table}[ht]
\centering
\begin{tabular}{c S[table-format=1.3] S[table-format=2] S[table-format=2.1] S[table-format=1.2] S[table-format=1.1]}
\hline
\hline
$(\varphi_\mathrm{ext}, \theta_\mathrm{ext})$ & \text{$\omega_{ge} / 2\pi\,(\si{\giga\hertz})$} & \text{$\chi\,(\si{\mega\hertz})$} & \text{$T_1\,(\si{\micro\second})$} & \text{$T_{2\mathrm{R}}\,(\si{\micro\second})$} & \text{$T_{2\mathrm{E}}\,(\si{\micro\second})$} \\
\hline
$(0,0)$ & 2.366 & 31 & 4.9 & 0.50 & 2.0 \\
$(\pi,0)$ & 0.792 & -12 & 40 & 0.48 & 2.4 \\
$(0,\pi)$ & 1.329 & -30 & 16 & 0.43 & 4.0 \\
$(\pi,\pi)$ & 1.460 & -33 & 9.2 & 0.33 & 1.9 \\
\hline
\hline
\end{tabular}
\caption{Coherence properties at the four distinct two-dimensional external flux sweet spots for the first excited state transition. Here, $\chi$ denotes the dispersive shift on the readout resonator, $T_1$ denotes the relaxation time, $T_{2\mathrm{R}}$ denotes the Ramsey decoherence time, and $T_{2\mathrm{E}}$ denotes the spin-echo decoherence time.\label{tab:coherence}}
\end{table}

\begin{figure*}
\centering
\includegraphics{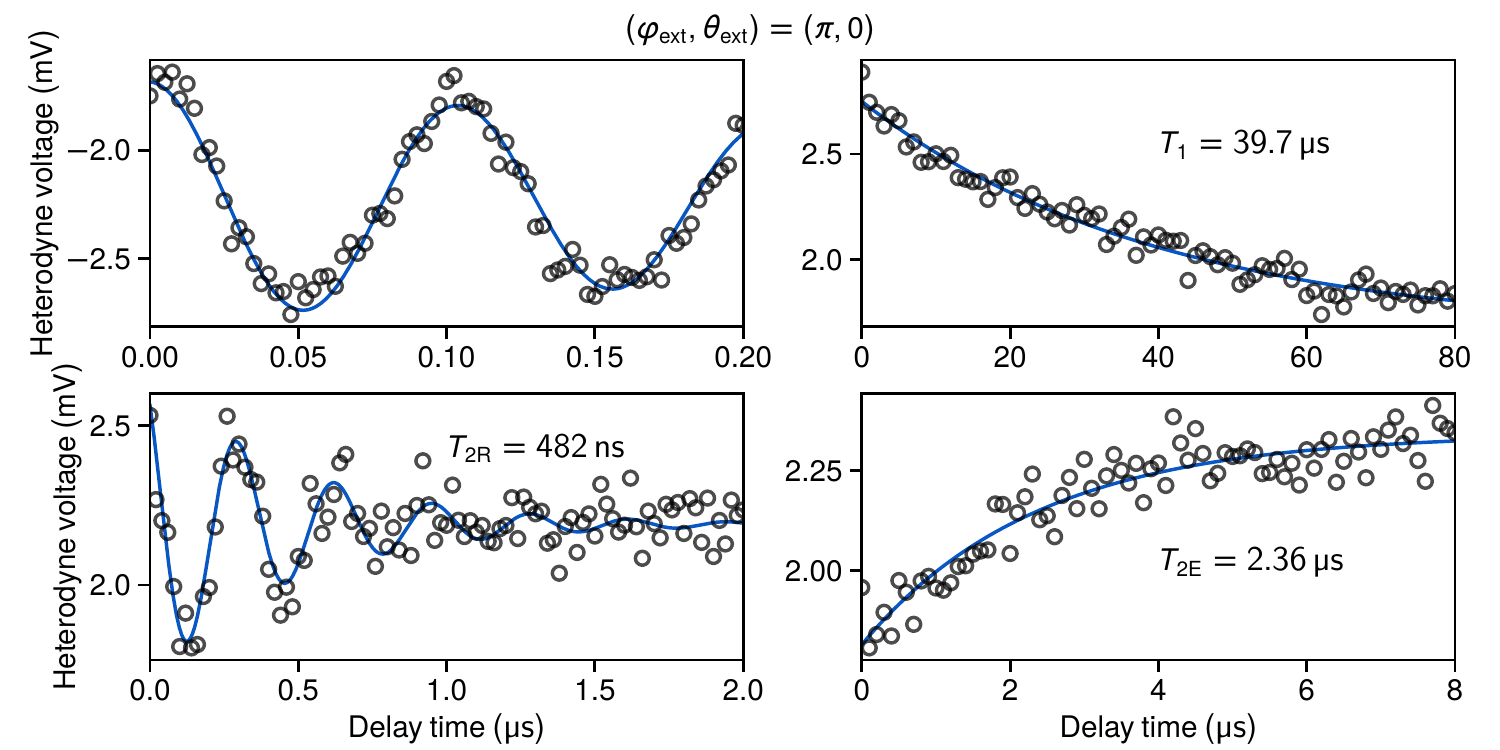}
\caption{Time-domain measurements of the first excited state transition at $(\varphi_\mathrm{ext}, \theta_\mathrm{ext}) = (\pi, 0)$. Clockwise from top left: Rabi oscillations, relaxation, spin-echo decoherence, and Ramsey decoherence.\label{fig:coherence}}
\end{figure*}

In addition to the spectroscopy data shown in Figs.\ \ref{fig:levelszero}--\ref{fig:levelshalf}, we also measured the coherence times and dispersive shifts on the readout resonator by first calibrating a $\pi$-pulse from measured Rabi oscillations (typical measurement and summary shown in Fig.\ \ref{fig:coherence} and Tab.\ \ref{tab:coherence}, respectively). The relatively large values of dispersive shift originate from the mutual inductance between the readout and circuit (shared junctions), which was designed to comprise a significant fraction of the readout resonator inductance. The relaxation times essentially follow a trend based on the transition frequency, suggesting that they are limited by dielectric loss. The small coherence times are likely due to flux noise exaggerated by the large loop sizes used in the circuit (see Fig.\ \ref{fig:sample}). This is consistent with the significant improvement in coherence time when adding a single echo pulse as well as the measured $T_{2\mathrm{R}} < \SI{0.1}{\micro\second}$ away from the sweet spots. We note that, although the sensitivity with respect to deviations in $\varphi_\mathrm{ext}$ changes substantially at the four sweet spots, the residual sensitivity with respect to $\theta_\mathrm{ext}$ renders the overall susceptibility to flux noise roughly constant. Additionally, we suspect that phase slips in the array junctions might contribute to dephasing.

\bibliography{biblio}

\end{document}